\documentclass[structabstract]{aa}

\usepackage{graphicx}
\usepackage{amssymb}
\usepackage{mathabx}
\usepackage{txfonts}
\usepackage{longtable,lscape}
\usepackage{natbib}
\usepackage{color}

\newcommand*{\mysim}{\mathord{\sim}}

\def\ergscm{erg~s$^{-1}$~cm$^{-2}$}

\def\arcmin{\hbox{$^\prime$}}

\def\apj{ApJ}

\def\aap{A\&A}
\def\mnras{MNRAS}

\begin{document}
   \title{INTEGRAL/IBIS nine-year Galactic Hard X-Ray Survey\thanks{Based on observations with INTEGRAL, an ESA project with
instruments and science data centre funded by ESA member states
(especially the PI countries: Denmark, France, Germany, Italy,
Switzerland, Spain), Czech Republic, and Poland, and with the
participation of Russia and the USA}}


   \author{R. Krivonos\inst{1,2},
          S. Tsygankov\inst{3,4,2},
          A. Lutovinov\inst{2},
          M. Revnivtsev\inst{2},
          E. Churazov\inst{1,2},
          R. Sunyaev\inst{1,2}
          }

   \institute{Max-Planck-Institut f\"ur Astrophysik,
              Karl-Schwarzschild-Str. 1, D-85740 Garching bei M\"unchen,
              Germany
         \and
         Space Research Institute, Russian Academy of Sciences,
         Profsoyuznaya 84/32, 117997 Moscow, Russia
         \and
         Finnish Centre for Astronomy with ESO (FINCA), University of Turku,  V\"ais\"al\"antie 20, FI-21500 Piikki\"o, Finland
         \and
         Astronomy Division, Department of Physics, FI-90014 University of Oulu, Finland
             }



  \abstract
   {The INTEGRAL observatory operating in a hard X-ray/gamma domain has gathered
     a large observational data set over nine years starting in 2003. Most of the observing time was dedicated to the Galactic source
     population study, making possible the deepest Galactic
     survey in hard X-rays ever compiled.}
   {We aim to perform a Galactic survey that can be used as the basis of
     Galactic source population studies, and perform mapping of the
     Milky Way in hard X-rays over the maximum exposure available at
     $|b|<17.5^{\circ}$.}
   {We used sky reconstruction algorithms especially developed for the high
     quality imaging of INTEGRAL/IBIS data.}
{We present sky images, sensitivity
   maps, and catalogs of detected sources in the three energy bands
   $17-60$, $17-35$, and $35-80$~keV in the Galactic plane at
   $|b|<17.5^{\circ}$. The total number of sources in the reference
   $17-60$~keV band includes 402 objects exceeding a $4.7\sigma$
    detection threshold on the nine-year time-averaged
   map. Among the identified sources with known and tentatively identified natures,
   253 are Galactic objects (108 low-mass X-ray binaries, 82 high-mass X-ray
   binaries, 36 cataclysmic variables, and 27 are of other types), and 115
   are extragalactic objects, including 112 active galactic nuclei (AGNs) and
   3 galaxy clusters. The sample of Galactic sources with
   $S/N>4.7\sigma$ has an identification completeness of
   $\sim92\%$, which is valuable for population studies. Since the survey is based on the nine-year sky maps, it is optimized for persistent sources and may be biased against finding transients.}
   {}


\keywords{Surveys -- X-rays: general -- Catalogs}
\authorrunning{Krivonos et al., }
\titlerunning{INTEGRAL 9-year Galactic Survey}
\maketitle

%

\section{Introduction}

A large fraction of astrophysical phenomena cannot be studied via
observations of individual sources, but require instead large
statistical studies. The last few decades have provided us with great
opportunities for studies of the populations of compact sources (black
holes, neutron stars, white dwarfs) in our Galaxy and nearby galaxies.

In particular, surveys of the sky in hard X-rays were performed with
the IBIS telescope \citep{ibis} of the INTEGRAL observatory
\citep{integral} and Burst Alert Telescope \citep[BAT;][]{bat} at the
\textit{Swift} observatory \citep{swift}. In contrast to
\textit{Swift}, with a nearly uniform all-sky survey, which is especially useful for
studies of active galactic nuclei
\citep[AGN;][]{2010ApJS..186..378T,palermo36,ajello12}, the INTEGRAL
observatory provides a sky survey with exposure that are deeper in the Galactic plane (GP) and has higher angular
resolution, which is essential in these crowded regions. This makes
the \textit{Swift}/BAT and INTEGRAL/IBIS surveys complementary to each
other.

The INTEGRAL observatory has been successfully operating in orbit
since its launch in 2002. Over the past few years, INTEGRAL data has
allowed us to construct high quality catalogs
\citep{2004AstL...30..382R,2006AstL...32..145R,2004AstL...30..534M,krietal05,krietal07b,kri2010b,2006ApJ...636..765B,2007ApJS..170..175B,2010ApJS..186....1B},
to reveal new types of sources
\citep{2003IAUC.8063....3C,revetal03a,2006ApJ...646..452S}, to
calculate the statistics of active galactic nuclei
\citep{2007A&A...462...57S,2006ApJ...636L..65B,beckmann09}, of low
mass X-ray binaries \citep{revetal08a}, of high mass X-ray binaries
\citep{2005A&A...444..821L,2007ESASP.622..241L,bodaghee07,bodaghee12},
and of cataclysmic variables \citep{revetal08b,scaringi10}.

In our previous papers \citep{kri2010a,kri2010b}, we presented the
seven-year hard X-ray all-sky survey in the energy range $17-60$~keV
based on the improved sky reconstruction method for the IBIS
telescope. The sensitivity of the survey was significantly improved by
suppressing the systematic noise.

Here we present the selected nine-year averaged sky images,
sensitivity maps, and catalog of the sources detected in the Galactic
plane ($|b|<17.5^{\circ}$) in three energy bands: $17-60$, $17-35$,
and $35-80$~keV. This survey is the most sensitive X-ray survey of the
Galaxy existing so far and it can be used as: 1) a basis for
statistical studies of different types of sources in our Galaxy, and
2) a guiding line for new surveys of a new generation of hard X-ray
focusing telescopes (e.g. NuSTAR described in \cite{nustar}, and
Astro-H in \cite{astro-h}).

The full set of sky maps is available at the SkyView Virtual
Observatory\footnote{http://skyview.gsfc.nasa.gov} \citep{skyview}
and Russian Science Data
Center\footnote{http://hea.iki.rssi.ru/integral} for the INTEGRAL
observatory at the Space Research Institute (IKI), Moscow.

\section{Survey}

To conduct the current Galactic survey, we selected publicly
available INTEGRAL data from December 2002 to January 2011 (spacecraft
revolutions 26-1013). Every individual INTEGRAL observation with
typical exposure time of $2$~ks (so called \textit{Science Window},
\textit{ScW}) was analyzed with a specially developed software package
(see e.g. \citealt{kri2010a} and references therein) to produce sky
images in three energy bands: $17-60$, $17-35$, and $35-80$~keV. In
contrast to our previous surveys, the flux scale in each \textit{ScW}
sky image was adjusted using the flux of the Crab nebula measured in
the nearest observations. This procedure was used to account for the
ongoing detector degradation and loss of sensitivity at low energies.

\begin{figure}
 \includegraphics[width=0.5\textwidth]{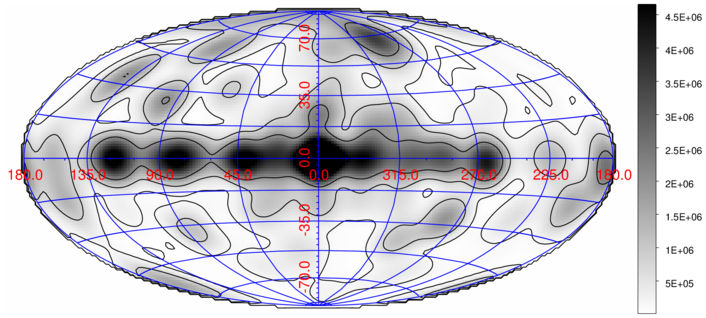}
\caption{Dead time-corrected exposure map of the INTEGRAL all-sky
  survey (January 2011, public data). The grid gap in the Galactic latitude is $17.5^{\circ}$, which is
  a half-height of the current Galactic survey. Blue contours represent
  exposure levels of $10$, $150$, $800$, $2000$, and $4000$~ks. The
  effective exposure in the GC region is $12$~Ms, which corresponds to
  $26$~Ms of a nominal time.}\label{fig:expomap}
\end{figure}

In total, we obtained 73489 sky images in each band, which comprises
$\sim132$~Ms of the effective (dead time-corrected) exposure. The
survey sky mapping was organized in six overlapping 
  $70^{\circ}\times35^{\circ}$ Galactic cartesian projections
centered on zero Galactic latitude ($b=0^{\circ}$) and
  $l=0^{\circ},\pm50^{\circ},\pm115^{\circ}$, and $l=180^{\circ}$.
The latitude coverage of the survey $|b|<17.5^{\circ}$ was chosen with
the IBIS/ISGRI field of view ($28^{\circ}\times28^{\circ}$) and
standard $5^{\circ}\times5^{\circ}$ observational pattern in
mind. Thus, we used all observations performed by INTEGRAL in the
Galactic plane. Fig.~\ref{fig:expomap} illustrates the INTEGRAL
exposure map of publicly available observations up to January 2011.

\begin{figure}[h]
 \includegraphics[width=0.5\textwidth]{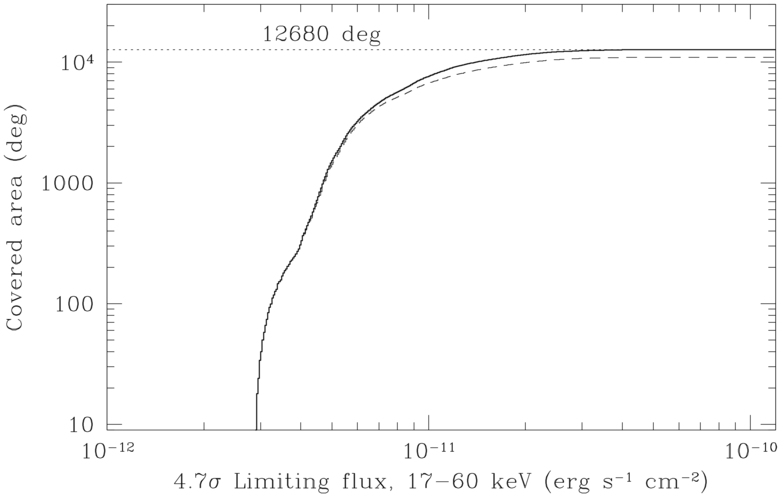}
\caption{Sky area as a function of the limiting flux for source
  detection with a $4.7\sigma$ significance (solid line). The black
  dashed curve shows the sky coverage with the masked area around
  bright sources listed in Table~\ref{tab:exclude}. The dotted line
  represents the geometrical area. }\label{fig:coverage}
\end{figure}

The survey sky coverage versus the $4.7\sigma$ limiting flux
is shown in Fig.~\ref{fig:coverage}. The peak sensitivity of the
survey is $2.9\times10^{-12}$~\ergscm ~($\sim0.20$~mCrab in
17-60~keV) at a $4.7\sigma$ detection level. The survey covers $90\%$
of the geometrical area ($12680$ degrees) down to the flux limit of
$2.0\times10^{-11}$~\ergscm ~($\sim1.41$~mCrab) and $10\%$ of the
total area down to the flux limit of $4.9\times10^{-12}$~\ergscm
~($\sim0.30$~mCrab).

\subsection{Systematic noise}

INTEGRAL/IBIS deep sky mosaics are usually affected by a systematic noise
caused by the source confusion in the region of GC and by the imperfect sky
reconstruction \citep{kri2010a}.

The large field of view (FOV) of the INTEGRAL/IBIS telescope leads to
a high probability of having some bright X-ray source within the
instrument FOV during any galactic observations. Therefore, to work at
the level of Poisson noise, the INTEGRAL/IBIS telescope
image reconstruction procedure should have a dynamic range of $10^3$
and more, which is very difficult to achieve owing to the imperfect
modelling of the mask shadow, the individual pixel sensitivity, the
variability of the background pattern, etc.

\begin{figure*}
  \centering
  \includegraphics[width=0.5\textwidth,angle=-90]{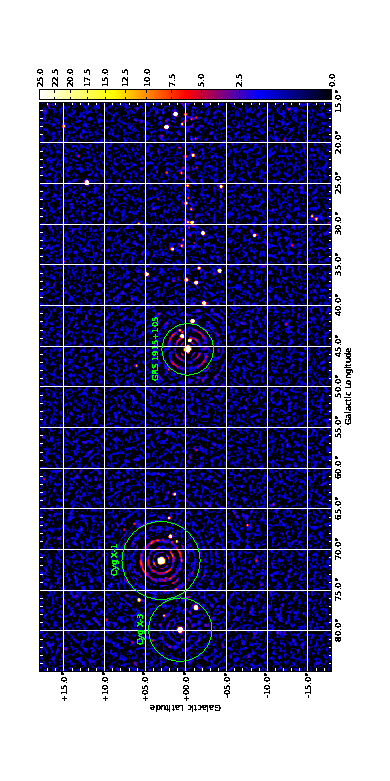}
  \caption{INTEGRAL/IBIS hard X-ray ($17-60$~keV) map of the sky
    region of Cyg X-1, Cyg X-3, and GRS 1915+105 at $l=+50^{\circ}$
    with masked area shown as green circles. The corresponding
    $S/N$ distribution of pixels is shown in
    Fig.~\ref{fig:skymap:histo}.}
  \label{fig:skymap}
\end{figure*}

\begin{figure}[h]
 \includegraphics[width=0.5\textwidth]{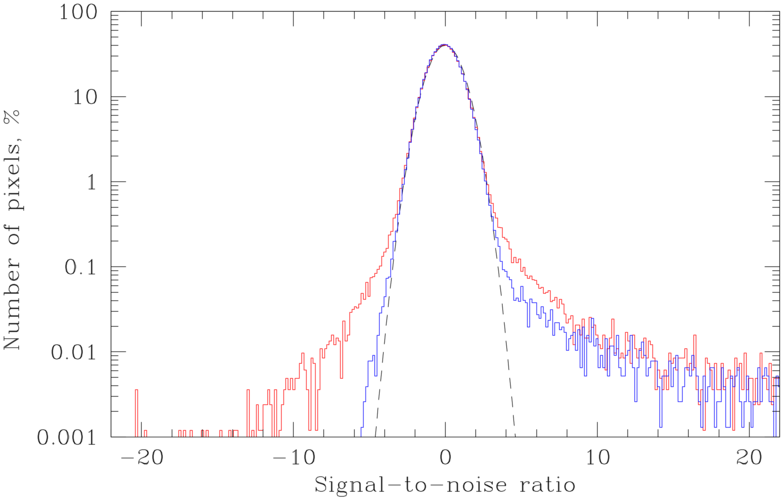}
\caption{Signal-to-noise ratio distribution of a number of pixels in
  the $17-60$~keV map of Fig.~\ref{fig:skymap}. Red and blue
  histograms show distributions of pixels in full and masked images,
  respectively. The dashed line represents the normal distribution
  with unit variance and zero mean.}\label{fig:skymap:histo}
\end{figure}

In spite of the latest version of the IBIS sky image reconstruction
allowing us to reach the dynamic range of the images $\sim10^3$, some
sky artefacts are still present around bright sources, such as Crab,
Sco X-1, Cyg X-1, Cyg X-3, Vela X-1, GX 301-2, and GRS 1915+105 (see a
sky mosaic at $l=+50^{\circ}$ in Fig.~\ref{fig:skymap}). To prevent
any false detections, we masked out circular regions around bright
sources with radii listed in Table~\ref{tab:exclude}. The exclusion
radius was chosen to contain all significant ($>4.7\sigma$) negative
excesses (indicators of the systematic noise, which is assumed to be
symmetric) around a given bright source. We note that known sources
with $S/N>10\sigma$ falling inside these regions were included in the
source catalog. A rejection of areas with high systematic noise
significantly improved the quality of the survey mosaics, which is
demonstrated in Fig.~\ref{fig:skymap:histo}, where we show a
distribution of signal-to-noise ($S/N$) values for pixels in the sky
mosaic at $l=+50^{\circ}$. As seen from Fig.~\ref{fig:skymap:histo},
the masked sky image does not contain strong deviations from the
Gaussian distribution, in contrast to the original one. Masked areas
around bright sources reduce the geometrical area of the survey by
13\% (Fig.~\ref{fig:coverage}).

\begin{figure*}
  \centering
  \includegraphics[width=0.5\textwidth,angle=-90]{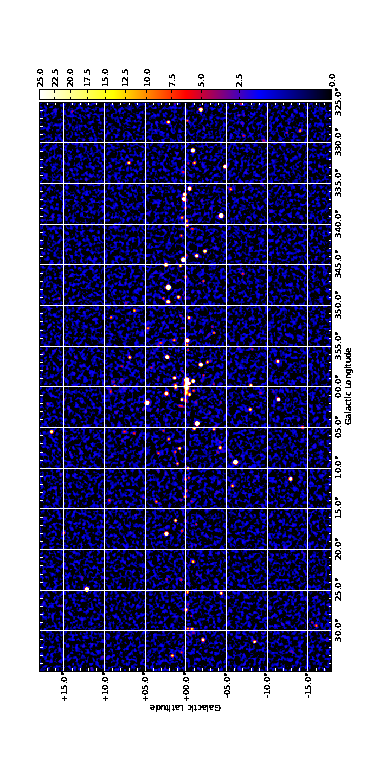}
  \caption{INTEGRAL/IBIS hard X-ray ($35-80$~keV) map of the sky
    region around the GC. The total exposure is about 26~Ms in the GC
    region. The image is shown in terms of $S/N$ with the color map
    ranging from values of 0 to 25. This color scheme is used to
    emphasize sky background variations. Fig.~\ref{fig:gcmap:histo}
    demonstrates a corresponding $S/N$ distribution of pixels.}
  \label{fig:gcmap}
\end{figure*}

\begin{figure}[h]
 \includegraphics[width=0.5\textwidth]{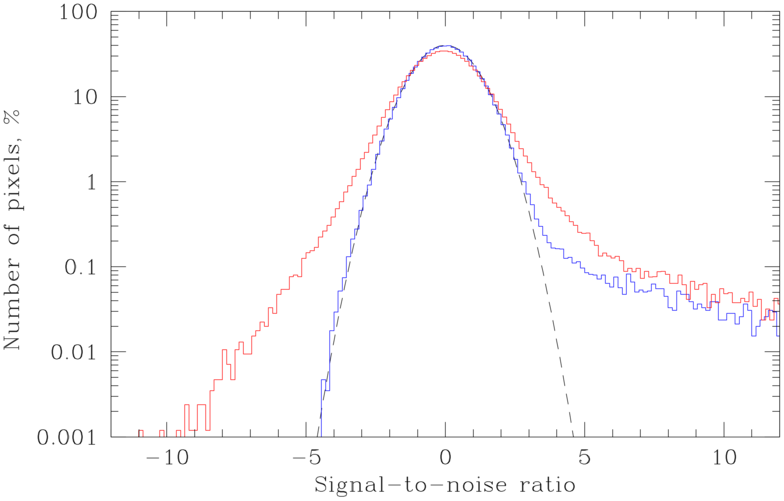}
\caption{Signal-to-noise ratio distribution of a number of pixels in
  the hard X-ray image shown in Fig.~\ref{fig:gcmap}. The dashed line
  represents the normal distribution with unit variance and zero
  mean. Red and blue histograms show pixel distributions for images in the 
  energy band $17-60$~keV (high systematic noise, see text) and
  $35-80$~keV, respectively.}\label{fig:gcmap:histo}
\end{figure}

Systematic effects are less important in the harder energy
bands. Fig.~\ref{fig:gcmap} shows that a sky image of the GC region in
the $35-80$~keV energy range is practically free of the systematic
noise with respect to $17-60$~keV band, which is confirmed by the
$S/N$ distribution of image pixels in Fig.~\ref{fig:gcmap:histo}.

\begin{table}
\caption{Exclusion radius around bright sources.}
\label{tab:exclude}
\centering
\begin{tabular}{l r r l }
\hline\hline
Name & Exclusion radius, deg. \\
\hline
Crab  &      23.2  \\
Sco X-1&     14.0  \\
Cyg X-1 &     4.8  \\
Cyg X-3 &     3.9  \\
Vela X-1 &    3.3  \\
GX 301-2&     2.2  \\
GRS 1915+105& 3.2  \\
\hline
\end{tabular}
\end{table}

\subsection{Detection threshold}

Given the IBIS/ISGRI angular resolution of the
survey, the sky map contains $\sim3\times 10^5$ statistically independent
pixels. Taking into account the minor contribution of the systematic noise, we adopted
a conservative detection threshold of $(S/N)_{\rm lim}>4.7\sigma$ to
ensure that the final catalog contains no more than one spurious
source assuming Poisson statistics.

A source detection in the region of $\mysim17$ degrees around GC
should be interpreted with special care because of the possible false
peaks induced by the systematic noise. The latter is revealed by a
distorted shape that differs significantly from the instrumental
point-spread function (PSF), which is a two-dimensional Gaussian
($\sigma=5$\arcmin). The sky map of the GC region contains $12$ peaks
above $4.7\sigma$ in the reference $17-60$~keV energy band. All these
candidate sources have a very distorted shape, and none of them have
been detected in the $35-80$~keV energy range, which also points to a
false detection. Therefore, all excesses in the GC region have been
attributed to the systematic noise.



\section{Catalog of sources}
\label{sec:catalog}

The catalog was compiled from the source sample exceeding the
detection threshold in the reference $17-60$~keV energy band.  The
list of sources is presented in Table~2, and its content is described
below.


{\it Column (1) ``Id''} -- source sequence number in the catalog.

{\it Column (2) ``Name''} -- source name. Their common names are given
for sources whose nature was known before their detection by
INTEGRAL. Sources discovered by INTEGRAL or those whose nature was
established thanks to INTEGRAL observations are named ``IGR''

{\it Columns (3,4) ``RA, Dec''} -- source equatorial (J2000) coordinates.

{\it Column (5,6,7) ``Flux''} -- time-averaged source flux in the
$17-60$~keV, $17-35$~keV, and $35-80$~keV energy bands, respectively.

{\it Column (8) ``Type''} -- general astrophysical type of the object:
LMXB (HMXB) -- low- (high-) mass X-ray binary, AGN -- active galactic
nucleus, SNR -- supernova remnant, CV -- cataclysmic variable, PSR
-- isolated pulsar or pulsar wind nebula (PWN), SGR -- soft gamma repeater,
RS CVn -- coronally active binary star, SymbStar -- symbiotic star,
Cluster -- cluster of galaxies. A question mark indicates that the
specified type is not firmly determined, so should be
confirmed. 

Determining of the natures of the sources is complicated and a
continually ongoing process. During the compilation of the catalog, we
selected from our point of view, the most recent or reliable,
identifications from the literature. In some cases (when references
are not given), the identifications had been performed using recent
observations of the Chandra, XMM-Newton, and Swift observatories, both
the Simbad\footnote{Simbad Astronomical Database
  http://simbad.u-strasbg.fr} and NED\footnote{NASA/IPAC Extragalactic
  Database http://ned.ipac.caltech.edu} database, as well as the
$2MASS$ catalog.

{\it Column (9) ``Ref.''} -- references. These are mainly provided for
new sources and are related to their discovery and/or nature. The
papers describing a routine analysis of a given source
e.g. confirmation of their nature, a refined position, etc. are not
referenced.

{\it Column (10) ``Notes''} -- additional notes such as type subclass, redshift
information, alternative source names. The redshifts of the
extragalactic sources were obtained from the Simbad and NED databases.

\onllongtabL{2}{
\begin{landscape}
\centering
\label{table:catalog}
\begin{longtable}{r|l|r|r|r|r|r|r|c|l|r|r|l}
\caption{The catalog of sources\tablefootmark{\dag} detected during the INTEGRAL/IBIS nine-year Galactic survey.}\\
\hline
\hline
Id & Name & RA & Dec & F$_{\rm 17-60~keV}$\tablefootmark{\star} & F$_{\rm 17-35~keV}$\tablefootmark{\star} & F$_{\rm 35-80~keV}$\tablefootmark{\star} & Type & Ref.\tablefootmark{\star\star} & Notes\tablefootmark{\star\star\star} \\
   &      & deg & deg & $\textrm{erg~} \textrm{cm}^{-2}\textrm{s}^{-1}$ & $\textrm{erg~} \textrm{cm}^{-2}\textrm{s}^{-1}$ & $\textrm{erg~} \textrm{cm}^{-2}\textrm{s}^{-1}$ &   &  & \\
\hline
\endfirsthead
\caption{Continued.} \\
\hline
Id & Name & RA\tablefootmark{\ddag} & Dec\tablefootmark{\ddag} & F$_{\rm 17-60~keV}$\tablefootmark{\star} & F$_{\rm 17-35~keV}$\tablefootmark{\star} & F$_{\rm 35-80~keV}$\tablefootmark{\star} & Type & Ref.\tablefootmark{\star\star} & Notes\tablefootmark{\star\star\star} \\
   &      & deg & deg & $\textrm{erg~} \textrm{cm}^{-2}\textrm{s}^{-1}$ & $\textrm{erg~} \textrm{cm}^{-2}\textrm{s}^{-1}$ & $\textrm{erg~} \textrm{cm}^{-2}\textrm{s}^{-1}$ &   &  & \\
\hline
\endhead
\hline
\endfoot
\hline
\endlastfoot
1 &  IGR J00040+7020  & 1.009 &  70.326  &  $0.85 \pm 0.12$ (7.4)  &  $0.50 \pm 0.08$ (6.5)  &  \textcolor{red}{$0.39 \pm 0.12$} (3.3)  &  AGN  & 1 & Sy2 z=0.096;\\ 
2 &  IGR J00234+6141  & 5.740 &  61.684  &  $0.73 \pm 0.09$ (8.5)  &  $0.55 \pm 0.06$ (9.6)  &  \textcolor{red}{$<0.18$}   &  CV  & 2,3 & \\ 
3 &  TYCHO SNR  & 6.332 &  64.140  &  $0.88 \pm 0.09$   &  $0.56 \pm 0.06$ (9.7)  &  $0.46 \pm 0.09$ (5.0)  &  SNR  &  & \\ 
4 &  SWIFT J0025.8+6818  & 6.374 &  68.357  &  $0.83 \pm 0.10$ (8.2)  &  $0.45 \pm 0.07$ (6.6)  &  $0.69 \pm 0.11$ (6.6)  &  AGN  &  & Sy2 z=0.012; $\dlsh$ \\
 & &  & &  &  &  & &  & $\drsh$  2MASX J00253292+6821442; \\
 & &  & &  &  &  & &  & $\drsh$  =IGR J00255+6821; \\
  
5 &  V709 Cas  & 7.204 &  59.292  &  $5.10 \pm 0.09$   &  $3.31 \pm 0.06$   &  $2.16 \pm 0.10$   &  CV  & 4 & \\ 
6 &  IGR J00291+5934  & 7.263 &  59.572  &  $2.33 \pm 0.09$   &  $1.27 \pm 0.06$   &  $1.58 \pm 0.09$   &  LMXB  & 5,6 & V* V1037 Cas\\
\end{longtable}
\tablefoot{ 
\tablefoottext{\dag}{For the column description see Sect.\ref{sec:catalog}.}
\tablefoottext{\ddag}{The positional accuracy of sources detected by IBIS/ISGRI depends on the source significance \citep{gros2003}. The estimated $68\%$ confidence intervals for sources detected at 5--6, 10, and $>20\sigma$ are $2.1$\arcmin, $1.5$\arcmin, and $<0.8$\arcmin,  respectively \citep{krietal07b}.}
\tablefoottext{\star}{Source flux was measured in the nine-year time-averaged map. The flux measured with $\sigma<4.7$ is highlighted in red. The flux measured at $S/N<2\sigma$ is shown as an $2\sigma$ upper limit in red. If a detection significance does not exceed $10\sigma$, it is shown in braces. Flux is expressed in units of $10^{-11}\textrm{erg}$~$\textrm{cm}^{-2}\textrm{s}^{-1}$.}
\tablefoottext{\star\star}{
(1)~\cite{2008A&A...482..113M},
(2)~\cite{2005ATel..394....1D},
(3)~\cite{2006AstL...32..588B},
(4)~\cite{2006MNRAS.372..224B},
(5)~\cite{2004ATel..352....1E},
(6)~\cite{2004ATel..353....1M},
(7)~\cite{2006ATel..939....1K},
(8)~\cite{2009A&A...495..121M},
(9)~\cite{2005A&A...433.1163D},
(10)~\cite{2004ATel..281....1D},
(11)~\cite{2010ApJS..186....1B},
(12)~\cite{2010A&A...519A..96M},
(13)~\cite{2005ATel..662....1K},
(14)~\cite{2009A&A...493..339W},
(15)~\cite{2010HEAD...11.1305B},
(16)~\cite{2012ATel.4151....1P},
(17)~\cite{2010ATel.2759....1L},
(18)~\cite{2012AstL...38....1L},
(19)~\cite{2006ATel..880....1B},
(20)~\cite{2008AstL...34..367B},
(21)~\cite{2011ATel.3382....1K},
(22)~\cite{2004A&A...413..309M},
(23)~\cite{1999MNRAS.306..100R},
(24)~\cite{2007ATel.1286....1T},
(25)~\cite{2008ApJ...674..686W},
(26)~\cite{2006ApJ...636..765B},
(27)~\cite{2005A&A...444L..37S},
(28)~\cite{2004ATel..261....1D},
(29)~\cite{2006A&A...459...21M},
(30)~\cite{2008A&A...487..509S},
(31)~\cite{2009AstL...35...33R},
(32)~\cite{2009A&A...502..787Z},
(33)~\cite{2006ATel..684....1K},
(34)~\cite{2006AstL...32..145R},
(35)~\cite{2006ATel..715....1M},
(36)~\cite{2006ATel..785....1M},
(37)~\cite{2006A&A...449.1139M},
(38)~\cite{2008A&A...477L..29L},
(39)~\cite{2010ApJ...720..987F},
(40)~\cite{2007ApJS..170..175B},
(41)~\cite{2004ATel..278....1P},
(42)~\cite{2008ApJ...685.1143T},
(43)~\cite{2004ATel..275....1G},
(44)~\cite{2010A&A...511A..48M},
(45)~\cite{2012AstL..Kar..subm},
(46)~\cite{2006ATel..783....1M},
(47)~\cite{2005ATel..519....1C},
(48)~\cite{2005ATel..528....1M},
(49)~\cite{2006A&A...446L..17B},
(50)~\cite{2005MNRAS.364..455C},
(51)~\cite{2009ApJ...701..811T},
(52)~\cite{2006ATel..810....1K},
(53)~\cite{2010ApJ...716..663R},
(54)~\cite{Extremesky2011Landi},
(55)~\cite{2012ATel.4183....1T},
(56)~\cite{2010MNRAS.408..975M},
(57)~\cite{2011ATel.3146....1M},
(58)~\cite{2007ATel.1034....1M},
(59)~\cite{2011ATel.3271....1L},
(60)~\cite{2004ATel..229....1W},
(61)~\cite{2006ApJ...647.1309T},
(62)~\cite{2007A&A...470..331M},
(63)~\cite{2005ATel..456....1S},
(64)~\cite{2005ApJ...631..506B},
(65)~\cite{2008ATel.1774....1K},
(66)~\cite{2010MNRAS.408.1866R},
(67)~\cite{2003AstL...29..587R},
(68)~\cite{2003IAUC.8063....3C},
(69)~\cite{2003IAUC.8076....1T},
(70)~\cite{2005A&A...433L..41L},
(71)~\cite{2003IAUC.8097....2R},
(72)~\cite{2010A&A...516A..94N},
(73)~\cite{2006A&A...447.1027B},
(74)~\cite{2004ATel..224....1T},
(75)~\cite{2006A&A...453..133W},
(76)~\cite{2004ATel..329....1L},
(77)~\cite{2005A&A...444..821L},
(78)~\cite{2003ATel..176....1M},
(79)~\cite{2008A&A...484..783C},
(80)~\cite{2005ATel..457....1G},
(81)~\cite{2008ATel.1396....1N},
(82)~\cite{2011MNRAS.411..137H},
(83)~\cite{2010AstL...36..904L},
(84)~\cite{2007ATel.1270....1B},
(85)~\cite{2011ApJ...731L...2K},
(86)~\cite{2009MNRAS.392L..11L},
(87)~\cite{2007A&A...467..529C},
(88)~\cite{2012ATel.4219....1D},
(89)~\cite{2003ATel..149....1K},
(90)~\cite{2003AstL...29..719L},
(91)~\cite{2005ATel..444....1G},
(92)~\cite{2007ESASP.622..373G},
(93)~\cite{2012A&A...538A.123M},
(94)~\cite{2005ApJ...634L..21B},
(95)~\cite{2006ATel..778....1B},
(96)~\cite{2005MNRAS.361..141G},
(97)~\cite{2004ATel..328....1L},
(98)~\cite{2010MNRAS.404.1591D},
(99)~\cite{2006ATel..874....1K},
(100)~\cite{2011ATel.3565....1G},
(101)~\cite{2012NewA...17..589N},
(102)~\cite{2003ATel..181....1S},
(103)~\cite{2006ApJ...638.1045S},
(104)~\cite{2004ATel..345....1K},
(105)~\cite{2009ApJ...701.1627H},
(106)~\cite{2004ATel..232....1B},
(107)~\cite{2004ATel..264....1T},
(108)~\cite{2011MNRAS.417L..26C},
(109)~\cite{2006ApJ...636..275B},
(110)~\cite{2007A&A...463..957K},
(111)~\cite{2003ATel..132....1R},
(112)~\cite{2006ATel..970....1B},
(113)~\cite{2006ATel..972....1W},
(114)~\cite{2005ATel..467....1G},
(115)~\cite{2008ATel.1445....1D},
(116)~\cite{2004A&A...425L..49R},
(117)~\cite{2009ATel.2132....1M},
(118)~\cite{2007AstL...33..149G},
(119)~\cite{2007ATel.1245....1K},
(120)~\cite{2010MNRAS.409L..69K},
(121)~\cite{2006ATel..885....1S},
(122)~\cite{2007ApJ...657L.109P},
(123)~\cite{2011MNRAS.411..620Z},
(124)~\cite{2004ATel..342....1G},
(125)~\cite{2007AstL...33..807C},
(126)~\cite{2003ATel..190....1S},
(127)~\cite{2005A&A...441L...1I},
(128)~\cite{2011A&A...533A...3C},
(129)~\cite{2003ATel..155....1L},
(130)~\cite{2007ATel.1054....1B},
(131)~\cite{2004AstL...30..382R},
(132)~\cite{2010A&A...510A..61T},
(133)~\cite{2004A&A...426L..41M},
(134)~\cite{2007A&A...474L...1N},
(135)~\cite{2005ApJ...629L.109U},
(136)~\cite{2007A&A...470..249F},
(137)~\cite{2007ATel.1248....1P},
(138)~\cite{2009ApJ...698..502B},
(139)~\cite{2008ATel.1686....1M},
(140)~\cite{2006ApJ...636L..65B},
(141)~\cite{2009MNRAS.395L...1B},
(142)~\cite{2003ATel..154....1L},
(143)~\cite{2005ApJ...630L.157M},
(144)~\cite{2004ATel..340....1R},
(145)~\cite{2008A&A...486..911N},
(146)~\cite{2003ATel..157....1C},
(147)~\cite{2004AstL...30..534M},
(148)~\cite{2008AIPC.1085..312T},
(149)~\cite{2012ATel.3951....1R},
(150)~\cite{2006AstL...32..221B},
(151)~\cite{2007AstL...33..159K},
(152)~\cite{2003IAUC.8088....4H},
(153)~\cite{2008ATel.1653....1S},
(154)~\cite{2006ATel..862....1T},
(155)~\cite{2007A&A...464..277M},
(156)~\cite{2006ATel..847....1H},
(157)~\cite{2008ATel.1623....1G},
(158)~\cite{2008AstL...34..653B},
(159)~\cite{2006A&A...455...11M},
(160)~\cite{2010MNRAS.403..945L},
(161)~\cite{2012ATel.4248....1M},
}
\tablefoottext{\star\star\star}{The spatial confusion with the source XXX is indicated as CXXX. The measured flux of sources being in spatial confusion should be taken with the caution.}
}
\end{landscape}
}

\section{Concluding remarks}

Our Galactic survey is based on sky maps averaged over a nine-year
time period, which obviously means that it has a strong bias against
finding low $S/N$ transient sources. This ensures that the current
survey contains mainly persistent objects, in addition to, however,
objects with strong intrinsic variability.

\begin{figure}[h]
 \includegraphics[height=0.5\textwidth,angle=-90]{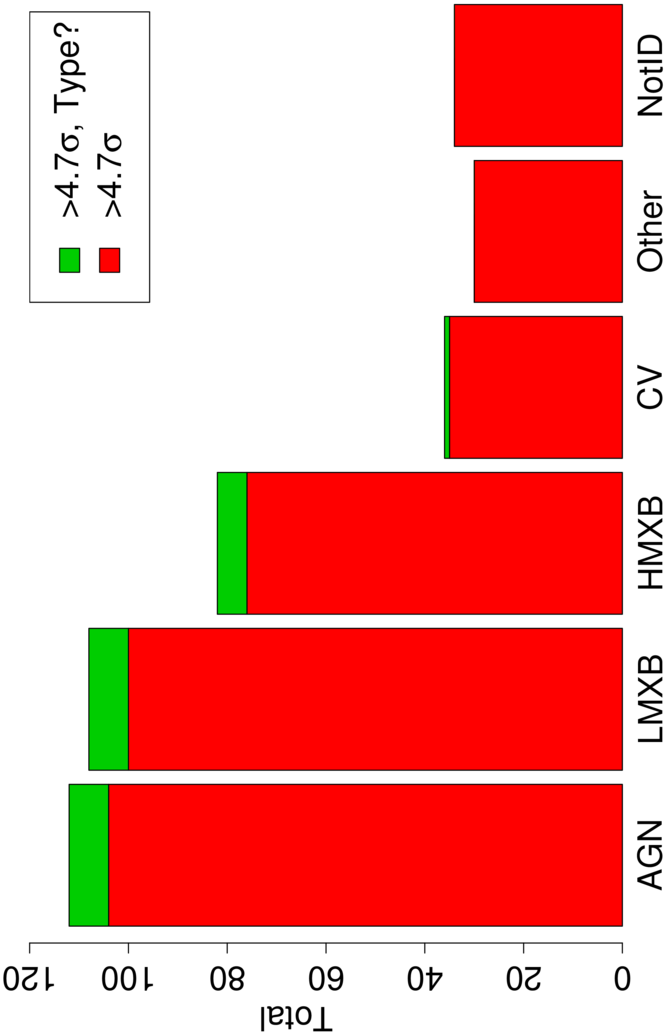}
\caption{Chart for the source classes detected at $S/N>4.7\sigma$ in the
  reference $17-60$~keV energy range (see
  Table~\ref{table:stat}). Green bar denotes the number of sources
  that have a tentative association with a given type of
  objects.}\label{fig:chart}
\end{figure}

We have presented detailed source statistics in Table~\ref{table:stat}
and a chart of source types in Fig.~\ref{fig:chart}. Our Galactic
survey has an identification completeness of
$(N_{Tot}-N_{NotID})/N_{Tot}=1-34/402\approx0.92$, which provides a
strong statistical basis for population studies. In the complementary
paper of Lutovinov et al., (2012, in prep.), we note in particular
that we used the source sample and sensitivity maps of the current
survey to study the HMXB luminosity function and their spatial
distribution in the Galactic disk.

\begin{table*}
\begin{minipage}[b]{\textwidth}\centering
\caption{Census of Galactic and extragalactic sources at $|b|<17.5^{\circ}$}
\begin{center}
\label{table:stat}
{\normalsize
\begin{tabular}{clllllll}
\hline
\hline
\noalign{\smallskip} Energy range, keV  & AGN & LMXB & HMXB & CV & Other &  NotID & Total \\
\noalign{\smallskip}\hline \noalign{\smallskip}
$17-60$ & 104+8$^{s}$  & 100+8$^{s}$ &  76+6$^{s}$ & 35+1$^s$  & 30  & 34   &  402 \\
$17-35$ &  89+7$^{s}$  &  98+7$^{s}$ & 74+6$^{s}$ & 35+1$^s$  & 28  & 22   &  367 \\
$35-80$ &  76+4$^{s}$  &  80+5$^{s}$ & 66+3$^{s}$ & 13        & 26  &  7   &  280 \\
\hline
\noalign{\smallskip}\hline\noalign{\smallskip}
\end{tabular}
}
\end{center}
\end{minipage}
\begin{minipage}[b]{\textwidth}
{\bf Notes.} The number of sources with a tentative classification is
denoted with $S$ index. Fig.~\ref{fig:chart} shows the chart
for source classes detected in the $17-60$~keV energy band.
\end{minipage}
\end{table*}

\begin{acknowledgements}
The data were obtained from the European and Russian INTEGRAL Science
Data
Centers\footnote{http://isdc.unige.ch}$^,$\footnote{http://hea.iki.rssi.ru/rsdc}.
This work was supported by the President of the Russian Federation
(through the program supporting leading scientific schools, project
NSH-5603.2012.2), by the Presidium of the Russian Academy of
Sciences/RAS (program P21 ``Non-stationary Phenomena in the
Universe''), grants 12-02-01265, 11-02-01328 and 11-02-12285-ofim-2011 from Russian
Foundation for Basic Research, State contract 14.740.11.0611. This
research has made use of the IGR Sources
page\footnote{http://irfu.cea.fr/Sap/IGR-Sources/} maintained by
J. Rodriguez \& A. Bodaghee.
\end{acknowledgements}

\end{document}